\documentclass[letterpaper, 10 pt, conference]{ieeeconf}

\IEEEoverridecommandlockouts                              

\usepackage{times}
\usepackage{amsmath}    
\usepackage{amssymb}    
\usepackage{mathrsfs}   
\usepackage{epsfig}     
\usepackage{graphicx}

\usepackage{comment}

\usepackage{subfigure}

\usepackage{amsfonts}
\usepackage{dsfont}

\usepackage{multirow}

\usepackage{cite}

\usepackage{enumerate}

\newcommand{\ep}{\epsilon}

\renewcommand{\H}{{\mathbf H}}

\newcommand{\E}{{\mathcal E}}
\newcommand{\V}{{\mathcal V}}
\renewcommand{\S}{{\mathcal S}}
\newcommand{\I}{{\mathcal I}}
\newcommand{\M}{{\mathcal M}}
\newcommand{\N}{{\mathcal N}}
\newcommand{\U}{{\mathbf U}}

\newcommand{\dE}{D_\Sigma}

\newcounter{constcount}
\newcommand{\eref}[1]{(\ref{#1})}

\newcounter{numcount}
\newcommand{\eqnum}{\stackrel{(\roman{numcount})}{=}\stepcounter{numcount}}

\newcommand{\cnt}{$(\roman{numcount})$ \stepcounter{numcount}}

\newcommand{\rescnt}{\setcounter{numcount}{1}}

\newcommand{\Nmimo}{\N_{\rm BC}}
\newcommand{\Hmimo}{\H_{\rm BC}}

\newcounter{thmcnt}
  \let\Oldsection\section
  
\renewcommand{\section}{\stepcounter{thmcnt}\Oldsection}

\allowdisplaybreaks

\newtheorem{theorem}{Theorem} 
\newtheorem{lemma}{Lemma} 
\newtheorem{definition}{Definition} 

\newtheorem{remark}{Remark} 
\newtheorem{question}{Question}[section]

\newcounter{examplecounter}
\newcommand{\aln}[1]{\begin{align*}#1\end{align*}}

\newcommand{\al}[1]{\begin{align}#1\end{align}}

\def\Item$#1${\item $\displaystyle#1$
   \hfill\refstepcounter{equation}(\theequation)}

\newcommand{\bea}{\begin{eqnarray}}
\newcommand{\eea}{\end{eqnarray}}
\newcommand{\beas}{\begin{eqnarray*}}
\newcommand{\eeas}{\end{eqnarray*}}

\begin{document}

\title{\LARGE \bf Informational Bottlenecks in Two-Unicast \\ Wireless Networks with Delayed CSIT}

\author{Alireza Vahid$^{1}$ \qquad Ilan Shomorony$^{2}$ \qquad Robert Calderbank$^{1}$
\thanks{$^{1}$Alireza Vahid and Robert Calderbank are with the department of Electrical and Computer Engineering, Duke University
        {\tt\footnotesize alireza.vahid@duke.edu} {\tt\footnotesize robert.calderbank@duke.edu}}%
\thanks{$^{2}$Ilan Shomorony is with the Department of Electrical Engineering and Computer Science, UC Berkeley
        {\tt\footnotesize ilan.shomorony@berkeley.edu}}%
}

 
\maketitle
\thispagestyle{empty}
\pagestyle{empty}

\begin{abstract}
We study the impact of delayed channel state information at the transmitters (CSIT) in two-unicast wireless networks with a layered topology and arbitrary connectivity.
We introduce a technique to obtain outer bounds to the degrees-of-freedom (DoF) region through the new graph-theoretic notion of \emph{bottleneck nodes}.
Such nodes act as informational bottlenecks only under the assumption of delayed CSIT, and imply asymmetric DoF bounds of the form $mD_1 + D_2 \leq m$.
Combining this outer-bound technique with new achievability schemes, we 
characterize the sum DoF of a class of two-unicast wireless networks,
which shows that, unlike in the case of instantaneous CSIT, the DoF of two-unicast networks with delayed CSIT can take an infinite set of values.
\end{abstract}



\section{Introduction}
\label{Section:Introduction}


Characterizing network capacity is one of the central problems in network information theory.
While a general solution is still far in the horizon, 
considerable research progress has been attained in several different fronts. 
In particular, single-flow networks 
are well understood and known to obey max-flow min-cute type principles, both for the case of wireline networks \cite{FF56, ACLY00} and for the case of wireless networks \cite{ADT11}.


When we consider multi-flow networks, however, the picture is much less clear.
As a natural first case to consider, networks with two source-destination pairs, or two-unicast networks, have recently been the focus of significant attention \cite{Ness2unicast,ShenviDey,xx,WangTwoUnicast,dof2unicastfull,zeng2015alignment},
but complete capacity characterizations are still a distant goal.
In fact, 
characterizing the capacity of two-unicast wireline networks is known to be as hard as the general $k$-unicast wireline problem \cite{TwoUnicastIsHard}.
In the wireless setting, matters become even more challenging since signals transmitted at different nodes interfere with each other, causing the two information flows to mix.

In an attempt to obtain first-order capacity approximations and capture the impact of interference in multi-flow wireless networks, a number of recent works have focused on characterizing the \emph{degrees of freedom} of different network configurations. 
Roughly speaking, the degrees of freedom (DoF) of a wireless network are the pre-log factor in the capacity expression, and can be thought of as the gain over time-sharing that can be obtained by carefully performing interference management and simultaneous routing of the information flows.
As a result of DoF studies, several new interference management techniques have recently been introduced, and shown to provide significant performance gains over simple time-sharing approaches \cite{CadambeJafar,MotahariRealInterference,xx,dofkkk}.
%
%
%
%
%
%
%
%
%
%
%
In particular, a careful combination of interference avoidance, interference neutralization, interference alignment \cite{CadambeJafar, MotahariRealInterference}, and aligned interference neutralization \cite{xx} was used in \cite{dof2unicastfull} to fully characterize the DoF of two-unicast layered wireless networks.

However, the promised gains in these DoF characterizations come at a high price.
In order to mitigate the effective interference experienced by the receivers, heavy coordination is required among all nodes in the network.
In particular, instantaneous channel state information (CSI) is assumed to be available at every transmitter.
%
%
In small-scale or slow-fading networks, the task of providing transmitters with CSI can be carried out with negligible overhead using feedback channels. 
However, as wireless networks grow in size, nodes turn mobile, and fast-fading channels become ubiquitous, 
providing up-to-date channel state information at the transmitters (CSIT) is practically infeasible, and CSIT is usually obtained with delay.
So how are the DoF gains promised by interference management techniques in multi-flow wireless networks affected by delayed CSIT?


%
%
%


Perhaps the first study of the impact of the delayed CSIT in wireless networks was~\cite{jolfaei1993new} where the delayed knowledge was used to create transmit signals that are simultaneously useful for multiple users in a broadcast channel. 
These ideas were then extended to different settings. 
Some examples are the study of erasure broadcast channels~\cite{georgiadis2009broadcast} and the capacity results for erasure interference channels~\cite{vahid2013capacity,vahid2013communication,FBBudgetISIT,vahid2015value}. In the context of multiple-input single-output (MISO) Gaussian broadcast channels (BC), it was shown that the delayed CSIT can still be very useful and in fact change the achievable DoF~\cite{Maddah-Tse-Allerton}.
This discovery 
generated a momentum in studying the DoF region of multi-antenna two-user Gaussian IC and X channel~\cite{GhasemiX1,Vaze_DCSIT_MIMO_BC,lashgari2014linear}, $k$-user Gaussian IC and X channel~\cite{Jafar_Retrospective,abdoli2011degrees}, and multi-antenna two-user Gaussian IC with delayed CSIT and Shannon feedback~\cite{tandon2012degrees,vaze2011degrees}. 

In this work, we study the impact of delayed CSIT in multi-hop multi-flow wireless networks by focusing our attention on two-unicast layered networks with arbitrary connectivity.
It is known that, in the case of instantaneous CSIT, the sum DoF of these networks can only take the values 1, 3/2 and 2, and can be determined based on two graph-theoretic structures \cite{ilanthesis}.
The first one is the notion of \emph{paths with manageable interference}, which captures when the two information flows can coexist and achieve a total of $2$ sum DoF.
The second one is the notion of an \emph{omniscient node}, which creates an \emph{informational bottleneck} and limits the DoF to $1$.
Whenever neither of these structures is found, $3/2$ DoF are achieavable.
The case of delayed CSIT was previously considered in \cite{WangTwoUnicastDCSIT}.
Interestingly, it was shown that as long as no omniscient node is found, at least $4/3$ DoF are achievable.
Hence, just as in the instantaneous CSIT case, the omniscient node is the key informational bottleneck whose absence determines when we can go beyond $1$ DoF, or simple time-sharing.
However, it is also known that unlike in the instantaneous CSIT case, networks with delayed CSIT may have $4/3$ DoF.
Two questions naturally arise:
How much richer is the set of possible DoF values in the delayed CSIT case? 
And what are the new informational bottleneck structures that apply only to the case of delayed CSIT?
 
%
%
%

In this paper, we make progress on both of these questions.
First, we generalize the concept of an omniscient node and introduce the notion of an \emph{$m$-bottleneck node}.
When a two-unicast network contains an $m$-bottleneck node for destination $d_1$, the DoF are constrained as 
$
m D_1 + D_2 \leq m,
$
where $D_i$ is the DoF for source-destination pair $i$.
Second, we show that, for $m \in \mathbb{N}$, there exists a two-unicast network with an $m$-bottleneck node where careful use of the delayed CSIT can achieve $(1-1/m,1)$ DoF, or $2-1/m$ sum DoF, matching the sum DoF outer bound implied by the $m$-bottleneck.
This establishes that, unlike in several recent DoF characterizations where the sum DoF are shown to only attain a small and finite set of values \cite{dof2unicastfull,XieTwoUnicastSecureDoF,WangX2Unicast}, the set of DoF values for two-unicast networks with delayed CSIT is in fact infinite.

%
%
%


\section{Problem Setting}
\label{Section:Problem}


A multi-unicast wireless (Gaussian) network $\N = (G,L)$ consists of a directed graph $G=(\V,\E)$, where $V$ is the node set and $\E \subset \V \times \V$ is the edge set, and a set of  source-destination pairs $L \subset \V \times \V$. 
We will focus on two-unicast Gaussian networks, which means that $L = \{ (s_1,d_1),(s_2,d_2)\}$, for distinct vertices $s_1, s_2, d_1,d_2 \in \V$. 
Moreover, we will assume that the network is \emph{layered}, meaning that the vertex set $V$ can be partitioned into $r$ subsets $\V_1,\V_2,...,\V_r$ (called layers) in such a way that $E \subset \bigcup_{i=1}^{r-1} \V_i \times \V_{i+1}$, and $\V_1 = \{s_1,s_2\}$, $\V_r = \{d_1,d_2\}$. 
For a vertex $v \in \V_j$, we will let $\I(v) \triangleq \{u\in V_{j-1}\, :\, (u,v) \in E\}$ be the set of parent nodes of $v$.

A real-valued channel gain $h_{i,j}[t]$ is associated with each edge $(i,j) \in E$ at each time $t$. 
We consider a fast-fading scenario, where the channel gains $\{h_{i,j}[t]\}_{t=0}^\infty$ for $i,j \in \V$ are assumed to be mutually independent i.i.d~random processes each obeying an absolutely continuous distribution with finite variance.
At time $t=1,2,...$, each node $i \in \V$ transmits a real-valued signal $X_{i}[t]$, which must satisfy an average power constraint $\frac1n \sum_{t=1}^n E\left[X_i^2[m]\right] \leq P$, $\forall \, v_i \in V$, for a communication block of length $n$. 
The signal received by node $j$ at time $t$ is given by
\al{ 
Y_j[t] = \sum_{i \in \I(j)} h_{i,j}[t] X_i[t] + Z_j[t], \label{receivedsignaldef}
}
where $Z_j[t]$ is the 
zero-mean unit-variance Gaussian noise at node $j$, assumed to be i.i.d.~across time and across nodes.
%
%
We will use $X_i^n$ to represent the vector $(X_i[0],...,X_i[n-1])$ and if $A$ is a subset of the nodes, $X_A[t] = (X_i[t] : i\in A)$.

We consider a delayed CSIT model where instantantaneous CSI is only available at the receiver of a given channel, and is learned with a unit delay at other nodes.
More precisely, we assume that at time $t$, a node $k \in \V$ has knowledge of 
\aln{
\{ h_{i,k}[t] : i \in \I(k)\} \cup \{ h_{i,j}[m] : (i,j) \in \E, 1 \leq m \leq t-1\}.
}
We will use $\H^t = \left( h_{i,j}[m] : (i,j) \in \E, 1 \leq m \leq t \right)$ to denote the random vector corresponding to the channel state information up to time $t$.
We point out that other more restrictive delayed CSIT models where nodes learn channel gains with a longer delay, or with a delay that is proportional to how far a given channel is in the network \cite{WangTwoUnicastDCSIT} can be considered.
However, it is straightforward to see that, through an interleaving operation, such models can be reduced to the model here considered.

We will use standard definitions for a coding scheme, an achievable rate pair $(R_1,R_2)$, and the capacity region $C(P)$ of a network $\N$.
We say that the DoF pair $(D_1,D_2)$ is achievable if we can find achievable rate pairs $(R_1(P),R_2(P))$ such that 
\aln{
D_i = \lim_{P\rightarrow \infty}\frac{R_i(P)}{\frac12 \log P}.
}
The sum DoF $\dE$ is defined as the supremum of $D_1+D_2$ over achievable DoF pairs.

%
%

%
%
%
%
%
%
%
%


\section{Main Results}
\label{Section:Main}


Several recent works on the DoF characterization of multi-flow networks revealed a similar phenomenon: for (Lebesgue) almost all values of channel gains, the DoF are restricted to a small finite set of values.
In \cite{dof2unicastfull} for instance, it is shown that $\dE \in \{1,3/2,2\}$ for two-unicast layered networks.
When the \emph{secure} DoF of two-unicast are considered instead, \cite{XieTwoUnicastSecureDoF} showed that we must have $\dE \in \{0,2/3,1,3/2,2\}$.
In \cite{WangX2Unicast}, two-source two-destination networks with arbitrary traffic demands were instead considered, and the set of DoF values was shown to be $\{1,4/3,3/2,2\}$.
Finally, for the delayed CSIT setting considered in this paper, \cite{WangTwoUnicastDCSIT} showed that, if $\dE \ne 1$, then $\dE \geq 4/3$, suggesting that perhaps in this case, $\dE$ is also restricted to a small number of discrete values.

In this paper, we show that this is not the case.
In fact, we prove the following:
\begin{theorem} \label{mainthm}
There exist two-unicast layered networks with delayed CSIT and sum DoF taking any value in the set 
\al{ \label{Seq}
\S \triangleq \left\{ 2\left(1 - \frac{1}{k} \right) : k=1,2,... \right\}  \cup \left\{ 2 \right\}. 
}
\end{theorem}
\vspace{2mm}


Intuitively, the reason why the sum DoF of two-unicast wireless networks can take all values in $\S$ is the fact that the delayed CSIT setting creates new informational bottlenecks in the network.
In this work, we identify a class of such structures, which we term $m$-bottleneck nodes.
We defer the formal definition of an $m$-bottleneck node to Section~\ref{bottlesec}, but we describe its significance with an example.
\begin{figure}[t]
\centering
\includegraphics[height = 3cm]{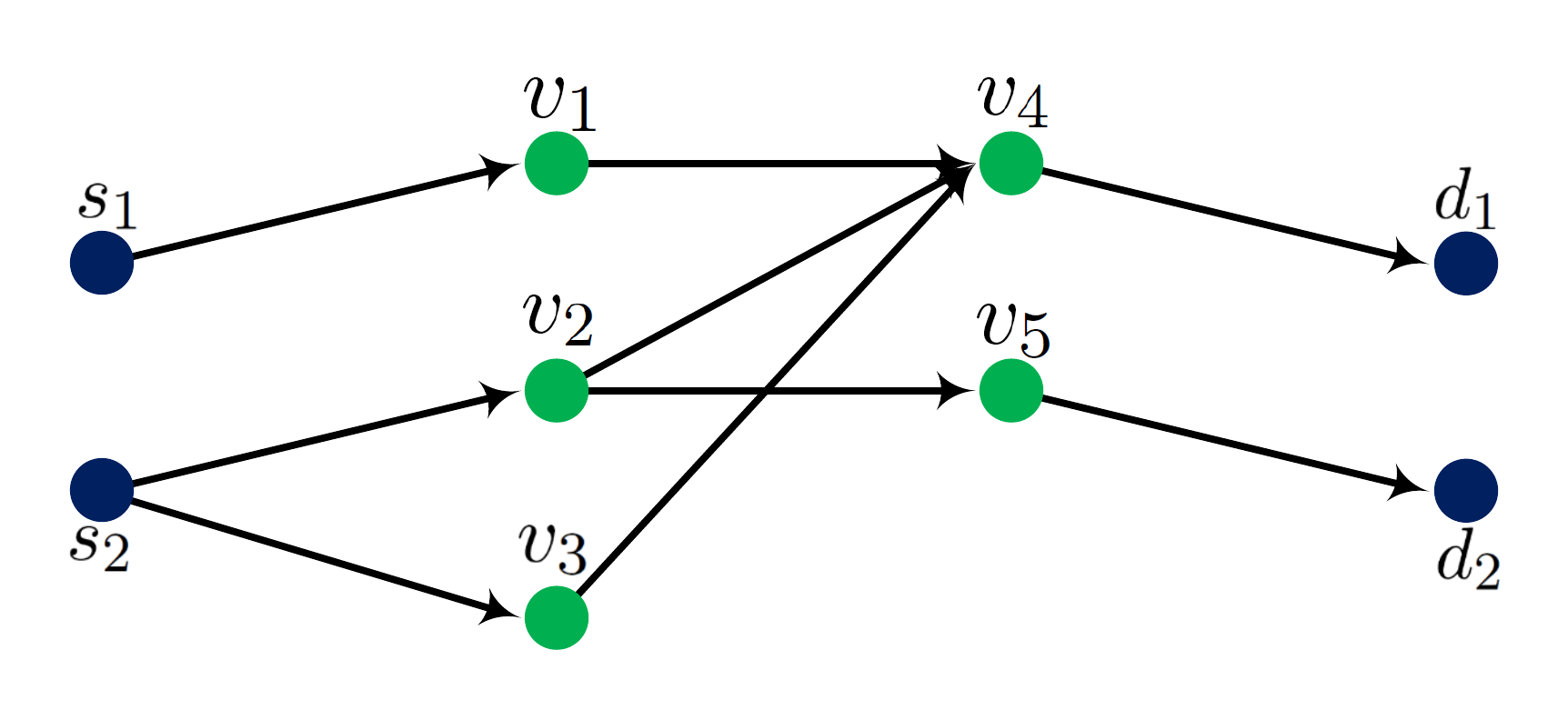}
\caption{Example of a network containing a bottleneck node ($v_4$).}\label{Fig:2D1D2}
\end{figure}
Consider the network in Fig.~\ref{Fig:2D1D2}.
If instantaneous CSIT were available, $v_2$ and $v_3$ could amplify-and-forward their received signals with carefully chosen coefficients so that their signals cancel each other at receiver $v_4$.
This would effectively create an interference-free network, and the cut-set bound of $2$ DoF would be achievable.
However, when only delayed CSIT is available, such an approach is no longer possible.
In fact, as we show in Section~\ref{bottlesec}, $v_4$ functions as a $2$-bottleneck node for destination $d_1$, causing the DoF to be constrained as
\aln{
2D_1 + D_2 \leq 2.
}
As it turns out, by utilizing delayed CSIT, the DoF pair $(1/2,1)$ can in fact be achieved.

In general, we show that whenever a network contains an $m$-bottleneck node for destination $d_i$ under the delayed CSIT assumption, we have
\al{ \label{bottlebound}
m D_{i} + D_{\bar{i}} \leq m,
}
where we let $\bar{i} = 3 - i$, and $i =1,2$. 
We point out that, in the case $m=1$, a bottleneck node reduces to the omniscient node \cite{WangTwoUnicast,ilanthesis,WangTwoUnicastDCSIT}, which was known to be an informational bottleneck in two-unicast networks, even under instantaneous CSIT.

In addition, we show that it is possible to build a two-unicast layered network where the outer bound implied by \eref{bottlebound} is tight.
In order to do so, we introduce \emph{linear} achievability schemes that make use of delayed CSIT in order to reduce the effective interference experienced by the bottleneck nodes as much as possible.
Theorem~\ref{mainthm} then follows by noticing that if we have a network with an $m$-bottleneck node for $d_1$ and an $m$-bottleneck node for $d_2$, then we must have $mD_1 + D_2 \leq m$ and $D_1+ m D_2 \leq m$, which implies
\aln{
(m+1) (D_1 + D_2) \leq 2m \; \Rightarrow \;  D_1 + D_2 \leq 2 - 2/(m+1).
}
Showing that two-unicast networks exist where the bound above is tight implies Theorem~\ref{mainthm}.
Before proving our main results, we present two motivating examples to describe the role of an $m$-bottleneck node.

\section{Motivating Examples}
\label{Section:Examples}


In this section, we first investigate the DoF of two networks, through which we motivate the idea of a bottleneck node and illustrate the transmission strategies that take advantage of delayed CSIT.


\subsection{Network with a bottleneck node} \label{bottleexsec}

\begin{figure}[t]
\centering
\includegraphics[height = 3.5cm]{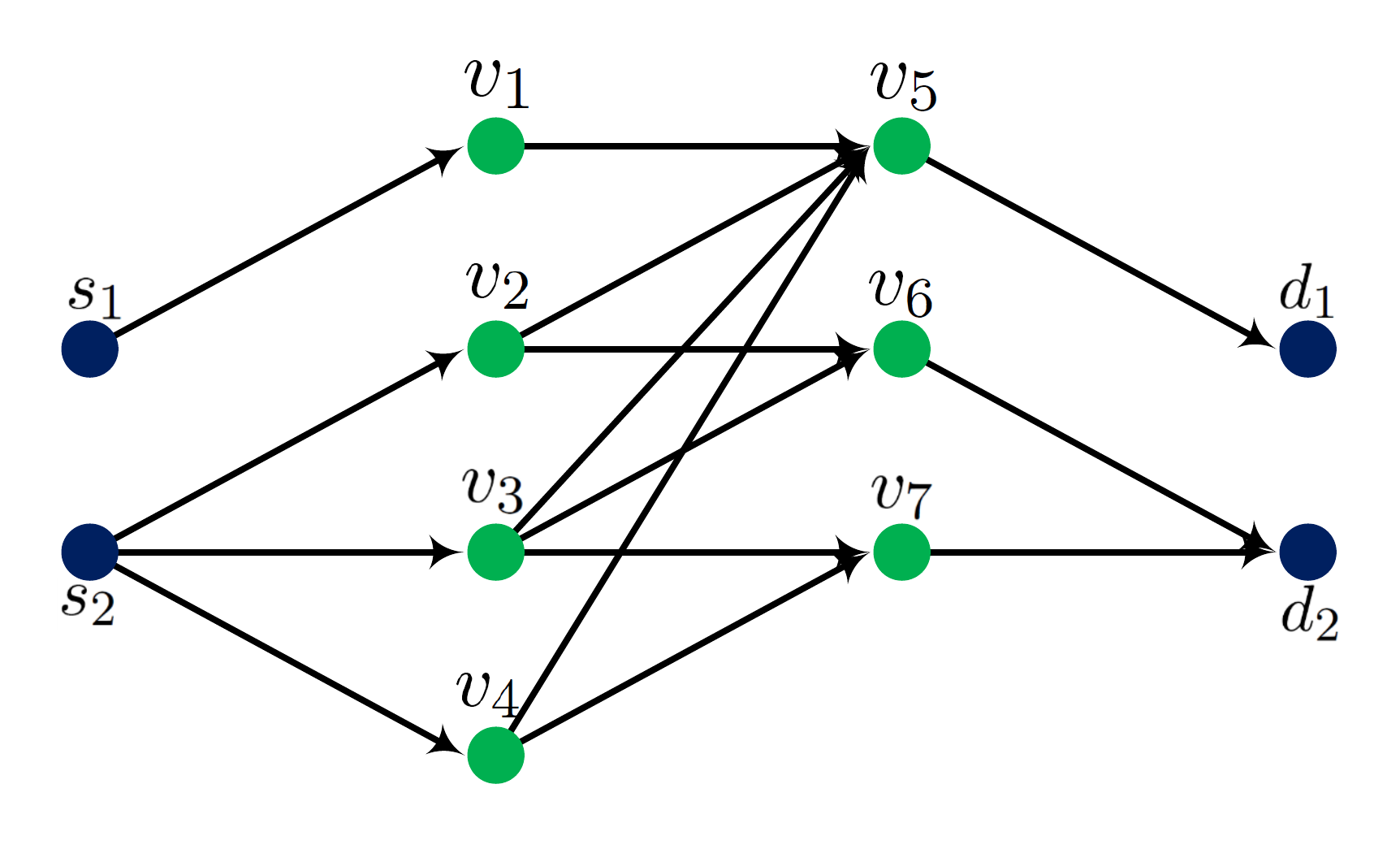}
\caption{Motivating example: we show that for this network, using the delayed CSIT, we can achieve $\left( D_1, D_2 \right) = \left( 2/3, 1 \right)$.}\label{Fig:3D1D2}
\vspace{-2mm}
\end{figure}

Consider the network depicted in Fig.~\ref{Fig:3D1D2}. 
If instantaneous CSIT was available, $v_2, v_3$ and $v_4$ could scale their signals according to using the information of $h_{v_2,v_5}$, $h_{v_3,v_5}$, and $h_{v_4,v_5}$ so that their interference at $v_5$ is canceled.
However, when CSIT is only available with a delay, such an approach does not work, and in order for information to flow from $s_2$ to $d_2$, some interference must inevitably occur at $v_5$.
This suggests that $v_5$ plays the role of a informational bottleneck, and the sum DoF should be strictly smaller than $2$.

In this subsection, we show that for this network we can achieve $\left( D_1, D_2 \right) = \left( 2/3, 1 \right)$. 
To do so, it suffices to show that during three time slots source $s_1$ can communicate two symbols to destination $d_1$, while source $s_2$ can communicate three symbols to destination $d_2$.
Since we can concatenate many three-slot communication blocks, we can describe our encoding as if the three time slots for the first hop occur first, followed by the three time slots for the second hop, and finally, the time slots for the third hop.
By concatenating many blocks, the delay from waiting three time slots at each layer becomes negligible. 
Next, we describe the transmission strategy for each hop separately. 
We will ignore noise terms to simplify the exposition.

\vspace{2mm}

\noindent {\bf Transmission strategy for the first hop:} During the first two time slots each source sends out two symbols; source $s_1$ sends out symbols $a_1$ and $a_2$, while source $s_2$ sends out symbols $b_1$ and $b_2$. During the third time slot, source $s_1$ remains silent while source $s_2$ sends out one symbol denoted by $b_3$. 
We note that upon completion of these three time slots, relay $v_1$ has access to symbols $a_1$ and $a_2$, and relay $v_j$ has access to symbols $b_1$, $b_2$, and $b_3$, $j=2,3,4$.

\vspace{2mm}

\begin{figure}[t]
\centering
\includegraphics[height = 4cm]{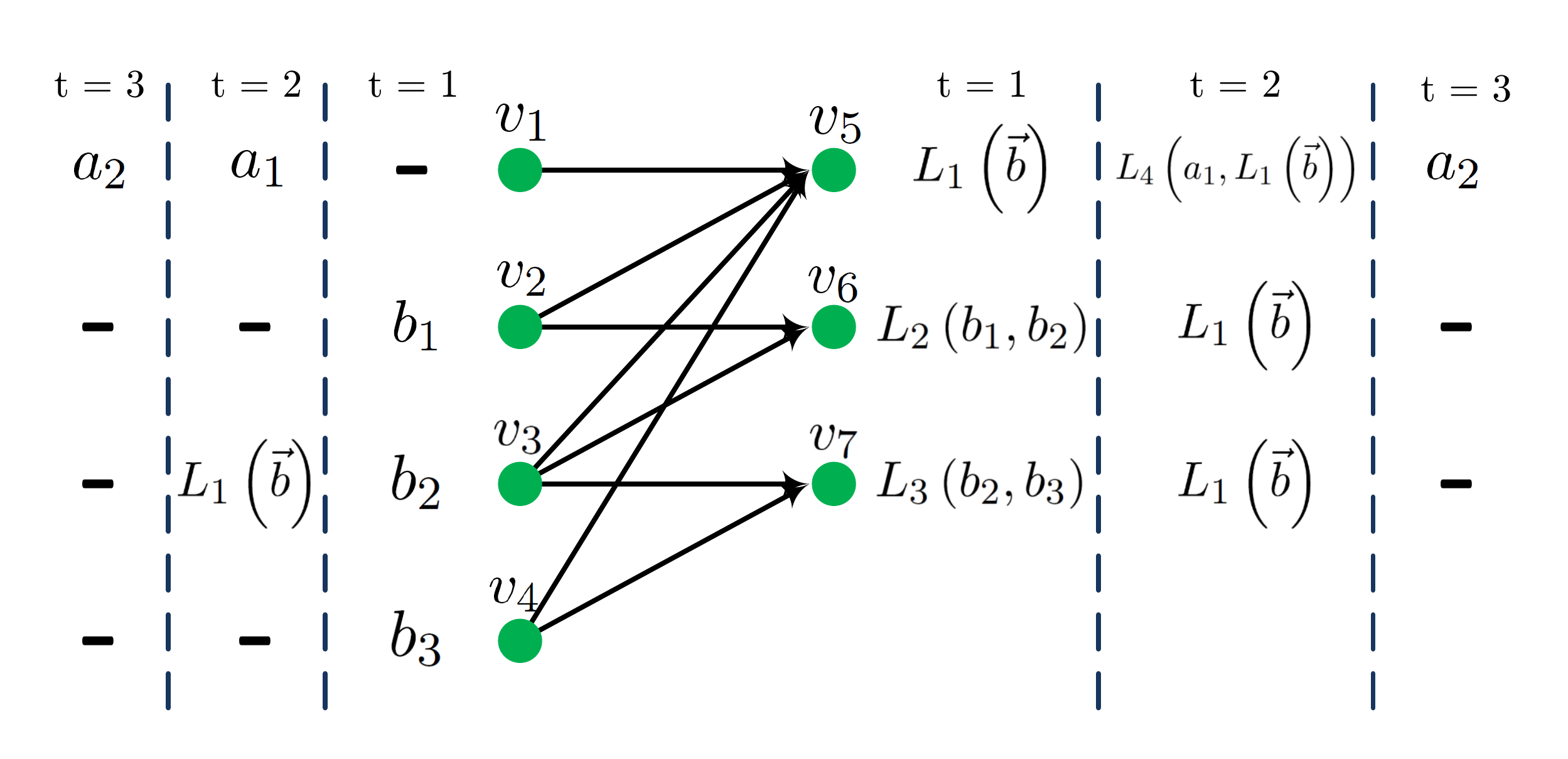}
\caption{Transmission strategy for the second hop of the network depicted in Fig.~\ref{Fig:3D1D2}.}\label{Fig:SecondHopEx1}
\vspace{-4mm}
\end{figure}

\noindent {\bf Transmission strategy for the second hop:} The key part of the transmission strategy happens in the second hop. During the first time slot, relay $v_2$ transmits $b_1$, relay $v_3$ transmits $b_2$, and relay $v_4$ transmits $b_3$ as depicted in Fig.~\ref{Fig:SecondHopEx1}. Ignoring the noise terms, relay $v_5$ obtains a linear combination of the symbols intended for destination $d_2$, $L_1\left( b_1,b_2,b_3 \right)$ that for simplicity we denote by $L_1(\vec{b})$. Similarly, relays $v_6$ and $v_7$ obtain linear combinations $L_2\left( b_1, b_2 \right)$ and $L_3\left( b_2, b_3 \right)$ respectively. During the first time slot, $v_1$ remains silent.

At this point, using the delayed knowledge of the channel state information, relay $v_3$ can (approximately) reconstruct $L_1(\vec{b})$. 
During the second time slot, relays $v_2$ and $v_4$ remain silent, relay $v_1$ sends out $a_1$, and relay $v_3$ sends out $L_1(\vec{b})$ (normalized to meet the power constraint). This way, $v_5$ obtains a linear combination of $a_1$ and $L_1(\vec{b})$ denoted by $L_4 (a_1, L_1(\vec{b}) )$. Note that $v_5$ already has access to $L_1(\vec{b})$ and thus can recover $a_1$. Also, note that $v_6$ and $v_7$ obtain $L_1(\vec{b})$.

Finally, during the third time slot, relays $v_2$, $v_3$ and $v_4$ remain silent, and relay $v_1$ sends out $a_2$. Upon completion of these three time slots, $v_5$ has access to $a_1$ and $a_2$, $v_6$ has access to $L_1(\vec{b})$ and $L_2\left( b_1, b_2 \right)$, and $v_7$ has access to $L_1(\vec{b})$ and $L_3\left( b_2, b_3 \right)$.


\vspace{2mm}
\noindent {\bf Transmission strategy for the third hop and decoding:} The transmission strategy for the third hop is rather straightforward.
Relay $v_5$ sends $a_1$ and $a_2$ to $d_1$, and relays $v_6$ and $v_7$ send three linearly independent equations $L_1(\vec{b})$, $L_2\left( b_1, b_2 \right)$, and $L_3\left( b_2, b_3 \right)$ to $d_2$.
%
%
%
%
Therefore $\left( D_1, D_2 \right) = \left( 2/3, 1 \right)$ DoF are achievable for the network of Fig.~\ref{Fig:3D1D2}.

\begin{figure}[ht]
\centering
\includegraphics[height = 4.4cm]{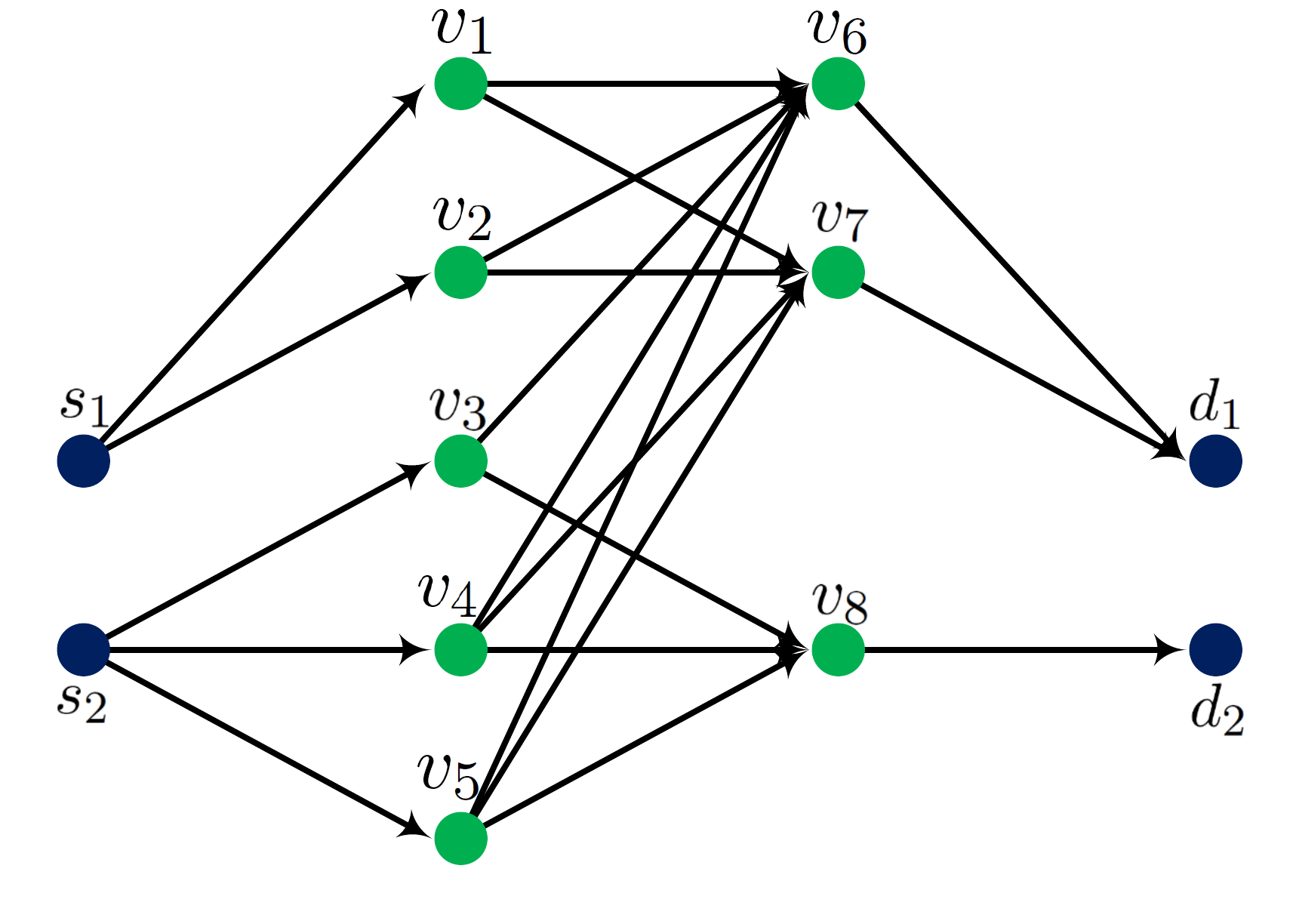}
\caption{In this example, we show we can achieve $\left( D_1, D_2 \right) = \left( 1, 1 \right)$.}\label{Fig:FullDoF}
\vspace{-2mm}
\end{figure}

\subsection{Network with no bottleneck node}

In this subsection, we consider the network in Fig.~\ref{Fig:FullDoF}. 
As in the previous example, the lack of instantaneous CSIT prevents nodes $v_3$, $v_4$ and $v_5$ from scaling their signals according to the channel gains of the second hop so that 
their interference at $v_6$ and $v_7$ is canceled.
Therefore, interference between the information flows is unavoidable.
However, as we will show, since there is no single node acting as a bottleneck node (as in the previous example), $\left( 1, 1 \right)$ DoF can be achieved. 
As it turns out, the diversity provided by an additional relay allows for a retroactive cancelation of the interference.

The transmission strategy has three time slots and the goal is for each source to communicate three symbols to its corresponding destination. 
For the first hop, the transmission strategy is very similar to that of the previous example and during each time slot, each source just sends a new symbol 
($a_i$'s for source $s_1$ and $b_i$'s for source $s_2$, for $i=1,2,3$).

\begin{figure}[t]
\centering
\includegraphics[width=\linewidth]{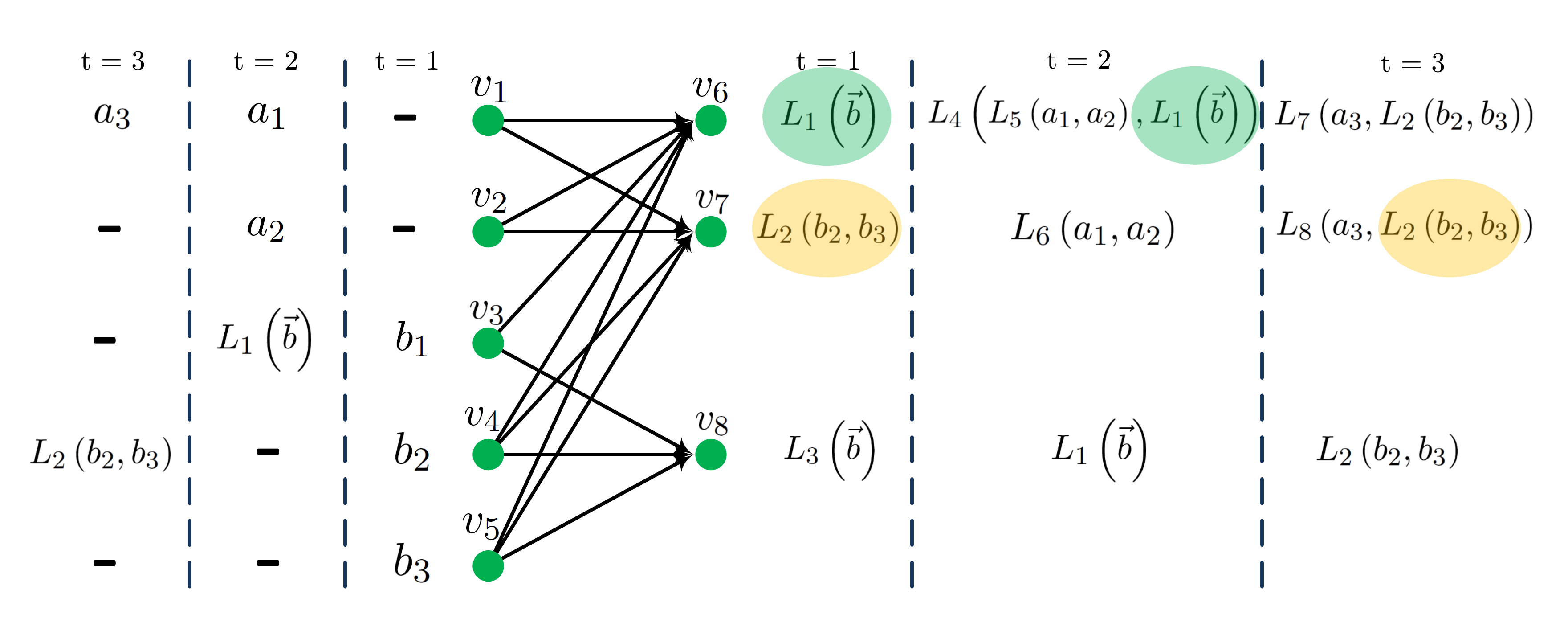}
\caption{Transmission strategy for the second hop of the network depicted in Fig.~\ref{Fig:FullDoF}.}\label{Fig:SecondHopEx2}
\vspace{-4.9mm}
\end{figure}

\vspace{2mm}
\noindent {\bf Transmission strategy for the second hop:} Similar to the previous example, the key part of the transmission strategy is in the second hop and that is what we focus on. The transmission strategy is illustrated in Fig.~\ref{Fig:SecondHopEx2} and described below.

During the first time slot, relays $v_1$ and $v_2$ remain silent. Relay $v_3$ sends out $b_1$, relay $v_4$ sends out $b_2$, and relay $v_5$ sends out $b_3$. 
Ignoring the noise terms, relay $v_6$ obtains a linear combination of all symbols intended for destination $d_2$ that we denote by $L_1 ( \vec{b} )$. Similarly, relay $v_7$ obtains $L_2\left( b_2, b_3 \right)$ and relay $v_8$ obtains $L_{3}(\vec{b})$.

At this point, using the delayed knowledge of the channel state information, relay $v_3$ can reconstruct $L_1(\vec{b})$ and relay $v_4$ can reconstruct $L_2\left( b_2, b_3 \right)$. During the second time slot, $v_3$ sends out $L_1(\vec{b})$ and this equation becomes available to relay $v_8$. During this time slot, relay $v_1$ sends out $a_1$ and relay $v_2$ sends out $a_2$. Note that due to the connectivity of the network, relay $v_6$ receives $L_4\left(  L_5\left( a_1, a_2 \right), L_1(\vec{b}) \right)$, and relay $v_7$ receives $L_6\left( a_1, a_2 \right)$. Using the received signals during the first two time slots, relay $v_6$ can recover $L_5\left( a_1, a_2 \right)$. Relays $v_4$ and $v_5$ remain silent during the second time slot.

In the third time slot, relay $v_1$ sends out $a_3$, and relay $v_4$ sends out $L_2\left( b_2, b_3 \right)$. All other relays remain silent. This way, relays $v_6$ and $v_7$ obtain $L_{7}\left( a_3, L_2\left( b_2, b_3 \right) \right)$ and $L_{8}\left( a_3, L_2\left( b_2, b_3 \right) \right)$ respectively. Now note that using the received signal during time slots one and two, relay $v_7$ can recover $a_3$.

\vspace{2mm}
\noindent {\bf Transmission strategy for the third hop and decoding:} In the third hop, relays $v_6$ and $v_7$ can easily communicate $L_5\left( a_1, a_2 \right)$, $L_{6}\left( a_1, a_2 \right)$, and $a_3$ to destination $d_1$ during the three time slots. Note that these equations are (with probability one) linearly independent, thus destination $d_1$ can recover its symbols. A similar story holds for destination $d_2$. This completes the achievability of $\left( D_1, D_2 \right) = \left( 1, 1 \right)$ for the network of Fig.~\ref{Fig:FullDoF}.


\section{Bottleneck Nodes}
\label{Section:Lemma}
\label{bottlesec}



As shown in the previous section, for the network of Fig.~\ref{Fig:FullDoF}, it is possible to exploit the diversity provided by the relays to retroactively cancel out the interference caused by relays $v_3$, $v_4$, and $v_5$ at relays $v_6$ and $v_7$. 
However, it is not difficult to see that the same approach cannot work for the network in Fig.~\ref{Fig:3D1D2}.
This suggests that the network in Fig.~\ref{Fig:3D1D2} contains an informational bottleneck that is not present in the network in Fig.~\ref{Fig:FullDoF} and that restricts the sum DoF to be strictly less than $2$. 
\begin{figure}[ht]
\centering
\includegraphics[height = 4cm]{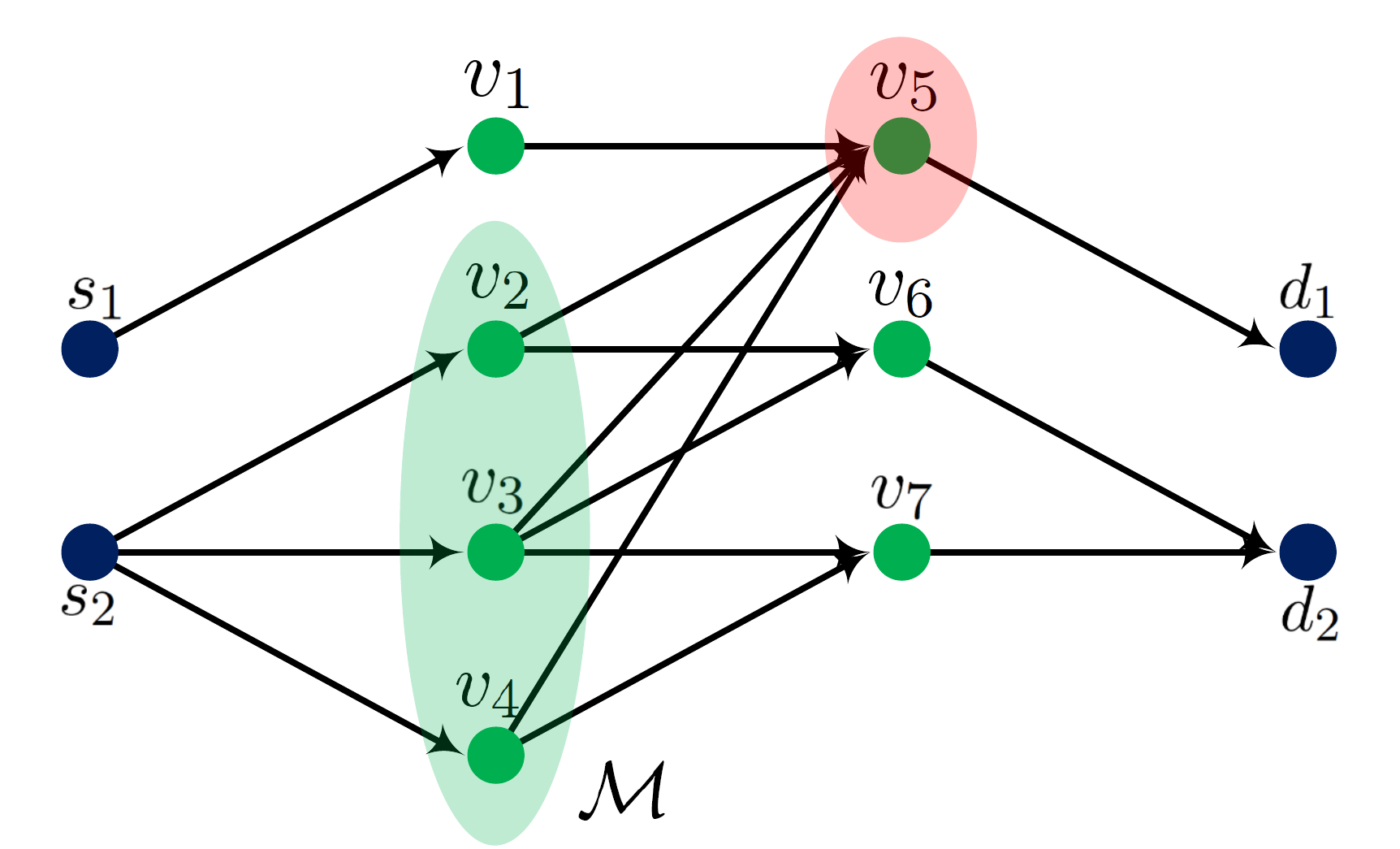}
\vspace{2mm}
\caption{$v_5$ acts as a bottleneck for the information flow. Relays $v_2$, $v_3$, and $v_4$ have to remain silent during a fraction of the time steps in order to allow $v_1$ and $v_5$ communicate. 
}
\label{Fig:3D1D2Bottleneck}
\end{figure}
As it turns out, this informational bottleneck is relay $v_5$.
Notice that the information flow from $s_1$ to $d_1$ must go through $v_5$.
Moreover, the fact that the information flow from $s_2$ to $d_2$ must go through the set of nodes $\M = \{v_2,v_3,v_4\}$, and CSIT is obtained with delay, makes interference between the flows unavoidable and
relays $v_2$, $v_3$, and $v_4$ have to remain silent during several time slots in order to allow $s_1$ and $d_1$ to communicate. 
As we will show in this section, the size of the set $\M$ determines how restrictive the bottleneck node $v_5$ is.
For the example in Fig.~\ref{Fig:3D1D2Bottleneck}, since $|\M| = 3$, the bottleneck node implies a bound of the form $3 D_1 + D_2 \leq 3$.

Before stating the main lemma on bottleneck nodes, we need a few definitions.






\begin{definition}
A set of nodes $A$, possibly a singleton, is a $(B,C)$-cut if the removal of $A$ from the network disconnects all paths from $B$ to $C$.
\end{definition}

\begin{definition}
A node $v$ is an omniscient node if it is an $(\{s_1,s_2\},d_i)$-cut and there is a node $u \in \I(v) \cup \{v\}$ that is a $(s_{\bar i},\{d_1,d_2\})$-cut.
\end{definition}

The existence of an omniscient node imposes that the sum DoF is bounded by $1$, even when instantaneous CSIT is available.
For more information regarding the omniscient node, we refer the readers to~\cite{ilanthesis,WangTwoUnicastDCSIT}. 
Motivated by the definition of an omniscient node, we introduce the notion of an $m$-bottleneck node, which reduces to an omniscient node in the case $m=1$.


\begin{definition} \label{bottledef}
A node $v \in \V$ is called an $m$-bottleneck node for $d_i$ if it is an $(\{s_1,s_2\},d_i)$-cut and there is a set $\M \subset \I(v)$ that is an $(s_{\bar i},\{d_1,d_2\})$-cut such that $\left| \M \right| = m$.
\end{definition}

Although a $1$-bottleneck node for $d_i$ is an omniscient node, the converse is not true.
The following theorem provides an outer-bound on the DoF of a two-unicast network with delayed CSIT and an $m$-bottleneck node for $d_i$.

\begin{theorem}
\label{bottlethm}
Suppose a layered two-unicast wireless network $\N$ contains an $m$-bottleneck node for $d_i$, for $i \in \{1,2\}$. Then under the delayed CSIT assumption, we have 
\begin{align} \label{thmeq}
m D_{i} + D_{\bar{i}} \leq m.
\end{align}
\end{theorem}

For $m=1$ the theorem follows since a $1$-bottleneck node for $d_1$ is an omniscient node. 
In the remainder of this section, we prove this result in the case $m>1$.
Suppose for network $\N$, we have a coding scheme that achieves $\left( D_1, D_2 \right)$ and  
$v$ is an $m$-bottleneck node for $d_1$ in layer $\V_{\ell+1}$. 
We use the network of Fig.~\ref{Fig:3D1D2Bottleneck}. 
In this network, it is straightforward to verify that node $v$ is a $3$-bottleneck node for $d_1$, according to Definition~\ref{bottledef}.

The proof contains two main steps, stated in two separate lemmas.
First, we construct a physically degraded MIMO BC, $\Nmimo$, where it is possible to achieve any DoF pair $(D_1,D_2)$ that is achievable in the original network $\N$.
Since the capacity of a physically degraded BC does not change with feedback, we can drop the delayed CSIT.
The second step is then to show that, if no CSIT is available, \eref{thmeq} must be satisfied in $\Nmimo$, which must therefore be satisfied in $\N$ as well.
We next describe these two steps in more detail.

We first construct the MIMO BC $\Nmimo$ based on $\N$ as follows.
The layer in $\N$ preceding the bottleneck node, $\V_\ell$, will become a sinlge source $s'$ with $|\V_\ell|$ antennas.
$\Nmimo$ will contain two receivers, namely $d_1^\prime$ and $d_2^\prime$. 


\begin{figure}[ht]
\centering
\includegraphics[height = 4cm]{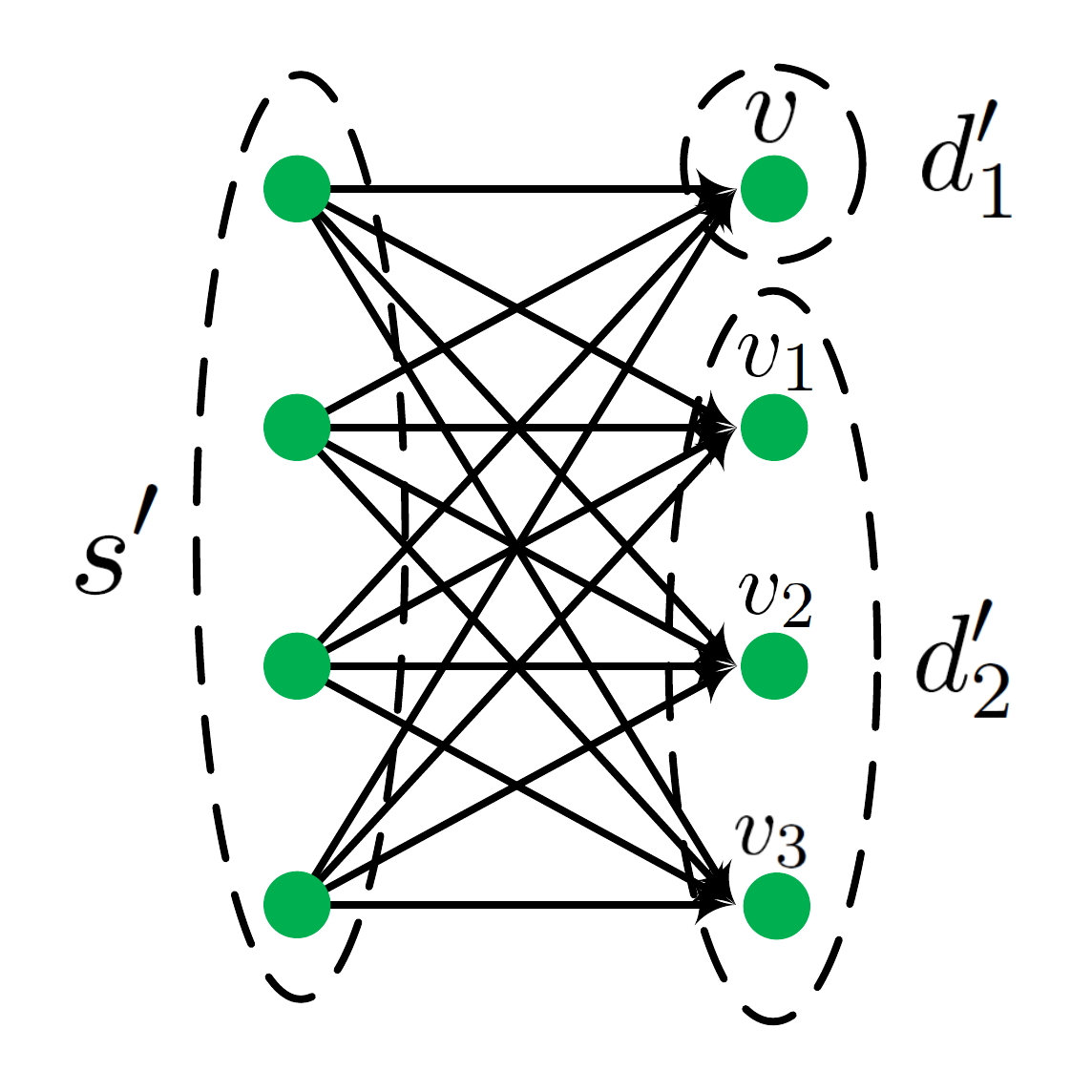}
\caption{Constructed MIMO BC $\Nmimo$ for the network $\N$ in Fig.~\ref{Fig:3D1D2Bottleneck}.}\label{Fig:3D1D2BottleneckProof3}
\end{figure}

Receiver $d_1^\prime$ has only one antenna, which is a replica of the bottleneck node $v$ in $\N$. 
On the other hand, receiver $d_2^\prime$ has $m$ receive antennas labeled as $v_1,v_2,\ldots,v_m$. See Fig.~\ref{Fig:3D1D2BottleneckProof3} for a depiction. 
The first receive antenna of $d_2^\prime$, $v_1$, has the same connectivity, channel realizations and noise realizations as that of node $v \in \N$. 
This guarantees that $Y_{v_1} = Y_v$, and that $\Nmimo$ is physically degraded.
The remaining antennas of $d_2^\prime$ each have statistically the same observation as $d_1^\prime \in \N^\prime$, but with independent channel and noise realizations.

\begin{lemma} \label{bclem1}
Any DoF pair $(D_1,D_2)$ achievable in $\N$ is also achievable in $\Nmimo$.
\end{lemma}

\begin{proof}
First we focus on the network $\N$, and assume we have a sequence of coding schemes that achieve a given rate pair $(R_1,R_2)$.
Since node $v$ is a bottleneck node for $d_1$, it is an $(\{s_1,s_2\},d_1)$-cut and must be able to decode $W_1$ as well, and we have 
\begin{align}
\label{eq:d1cut}
H\left( W_1|Y_{v}^n, \H^n \right) \leq n \epsilon_n,
\end{align}
where $\ep_n \to 0$ as $n \to \infty$, from Fano's inequality.
Next we notice that from the received signals in any given layer one should be able to reconstruct $W_2$, and we have 
\begin{align}
\label{eq:d2cut}
H & \left( W_2| \left[ F_{\M,\V_{\ell+1}} X_{\M} + Z_{\V_{\ell+1}} \right]^n, X_{\M^c}^n, \H^n \right) \nonumber \\
&  \leq H \left( W_2| Y_{\V_{\ell+1}}^n, \H^n \right)   \leq  n \epsilon_n,
\end{align}
where we let $F_{\M,\V_{\ell+1}}$ be the transfer matrix between $\M$ and $\V_{\ell+1}$, and $\M^c = \V_{\ell} \setminus \M$.
Our goal will be to emulate network $\N$ in the MIMO BC $\Nmimo$, so that destination $d_1'$ can recreate $Y_1^n$ to decode $W_1$, and destination $d_2'$ can approximately recreate $\left[ F_{\M,\V_{\ell+1}} X_{\M} + Z_{\V_{\ell+1}} \right]^n$ and $X_{\M^c}^n$ to decode $W_2$.

The main idea is to have the source $s'$ in $\Nmimo$ simulate all the layers in $\N$ up to $\V_\ell$.
In order to do that, let's first suppose that $s'$ and the destinations can share some randomness, drawn prior to the beginning of communication block.
This shared randomness corresponds to noise and channel realizations for the network $\N$ 
during a block of length $n$.
Let us denote these noise and channel realizations with the random vector $\U$.
Notice that the channel and noise realizations in $\U$ are independent of the actual channel and noise realizations in $\Nmimo$.
Using $\U$ and the messages $W_1$ and $W_2$, $s'$ can transmit what the nodes in layer $\V_\ell$ from $\N$ would have transmitted (same distribution).

Since the received signal at $d_1'$ has the same distribution as the received signal at $v$ in network $\N$, similar to \eref{eq:d1cut}, for $\Nmimo$, we have 
\begin{align}
\label{eq:d1cutmimo}
H\left( W_1|Y_{d_1'}^n, \Hmimo^n, \U \right) \leq n \epsilon_n.
\end{align}

Similarly, since the first antenna of $d_2'$ receives the exact same signal as $d_1'$, we have
\begin{align}
H & \left( W_2 |Y_{d_2'}^n, \Hmimo^n, \U \right) \nonumber \\
\leq H & \left( W_2 |W_1, Y_{d_2'}^n, \Hmimo^n, \U \right) + H\left( W_1 |Y_{d_2'}^n, \Hmimo^n, \U \right) \nonumber \\
\leq  H & \left( W_2 |W_1, Y_{d_2'}^n, \Hmimo^n, \U \right) + n \epsilon_n. \label{eq:cut2}
\end{align}

Next we notice that in $\N$, $X_{\M^c}$ is only a function of $\U$ and $W_1$. As a result, the source $s'$ in $\Nmimo$  can reconstruct $X_{\M^c}$ and transmit it from the corresponding antenna in $\Nmimo$.
This is because $\M$ is a $(s_{2},\{d_1,d_2\})$-cut in $\N$ and there can be no path from $s_2$ to $\M^c$.
Therefore, if we let $F_{\M,\V_{\ell+1}}$ be the transfer matrix drawn as part of the shared randomness $\U$, 
 and $\tilde Z_{\V_{\ell+1}}$ be a noise vector identically distributed as $Z_{\V_{\ell+1}}$ in $\N$ but independent from everything else, we have
\al{
& H \left( W_2|  W_1, Y_{d_2'}^n, \Hmimo^n, \U \right) \nonumber \\
& = H \left( W_2| \left[ F_{\M,\V_{\ell+1}} X_{\M} + \tilde Z_{\V_{\ell+1}} \right]^n, Y_{d_2'}^n, W_1, \Hmimo^n, \U \right)  \nonumber \\
& \; + I \left( W_2; \left[ F_{\M,\V_{\ell+1}} X_{\M} + \tilde Z_{\V_{\ell+1}} \right]^n | Y_{d_2'}^n, W_1, \Hmimo^n, \U \right) \nonumber \\
& \leq H \left( W_2| \left[ F_{\M,\V_{\ell+1}} X_{\M} + \tilde Z_{\V_{\ell+1}} \right]^n, X_{\M^c}^n, \U \right)  \nonumber \\
& \; + I \left( W_2; \left[ F_{\M,\V_{\ell+1}} X_{\M} + \tilde Z_{\V_{\ell+1}} \right]^n | Y_{d_2'}^n, W_1, \Hmimo^n, \U \right), \label{miterm}
}
and, from \eref{eq:d2cut}, the first term above is upper-bounded by $n\ep_n$.
All we need to show is that the mutual information term in \eref{miterm} is $o(\log P)$.
Let $G_{\M,d_2'}$ and $G_{\M^c,d_2'}$ be the transfer matrices from $\M$ and $\M^c$ to $d_2'$ in $\Nmimo$.
Notice that $G_{\M,d_2'}$ is an $m \times m$ matrix and is invertible with probability $1$.
Therefore, from $W_1$, $\Hmimo$, and $\U$ we can build $G_{\M^c,d_2'} X_{\M^c}$, and then use it to compute
\aln{
& F_{\M,\V_{\ell+1}} G_{\M,d_2'}^{-1} \left( Y_{d_2'} - G_{\M^c,d_2'} X_{\M^c} \right)  \\
& =  F_{\M,\V_{\ell+1}} X_{\M} + \hat Z,
}
where $\hat Z$ is a combination of noise terms, whose power is a function of channel gains, but not of $P$.
Therefore, the mutual information term in \eref{miterm} can be upper bounded as 
\al{
& h(\tilde Z_{\V_{\ell+1}}^n - \hat Z^n) \nonumber \\
& \quad - h\left( \left[ F_{\M,\V_{\ell+1}} X_{\M} + \tilde Z_{\V_{\ell+1}} \right]^n | Y_{d_2'}^n, W_1, W_2, \Hmimo^n, \U \right) \nonumber \\
& = h(\tilde Z_{\V_{\ell+1}}^n - \hat Z^n) - h\left( \tilde Z_{\V_{\ell+1}}^n | Y_{d_2'}^n, W_1, W_2, \Hmimo^n, \U \right) \nonumber \\
& = h(\tilde Z_{\V_{\ell+1}}^n - \hat Z^n) - h\left( \tilde Z_{\V_{\ell+1}}^n \right) \nonumber \\ 
& \leq n \, o(\log P), \label{ologpeq}
}
where the first equality follows since $\tilde F_{\M,\V_{\ell+1}} X_{\M}$ is just a function of $W_1$, $W_2$ and $\U$.
Therefore, from \eref{eq:cut2}, \eref{miterm} and \eref{ologpeq}, we have 
\aln{
H(W_2 | Y_{d_2'}^n, \Hmimo, \U) \leq n \ep_n + n \, o( \log P).
}
Hence, under the assumption of shared randomness, any pair $(D_1,D_2)$ achievable on $\N$ is also achievable in $\Nmimo$.
But since the shared randomness is drawn independently from $W_1$ and $W_2$, we can simply fix a value $\U = {\bf u}$ for which the resulting error probability is at most the error probability averaged over $\U$.
Thus, the assumption of shared randomness can be dropped, and the lemma follows.
%
%
%
%
%
%
\end{proof}

Lemma~\ref{bclem1} allows us to bound the DoF of network $\N$ by instead bounding the DoF of $\Nmimo$.

\begin{lemma} \label{bclem2}
For the MIMO BC $\Nmimo$, $m D_1 + D_2 \leq m$.
\end{lemma}

\begin{proof}
The MIMO BC $\Nmimo$ is physically degraded since the first antenna of $d_2^\prime$ observes the same signal as $d_1^\prime$. 
We know that for a physically degraded broadcast channel, feedback does not enlarge the capacity region \cite{ElGamal-Degraded}. 
Therefore, we can ignore the delayed knowledge of the channel state information at the transmitter (\emph{i.e.} no CSIT assumption). 
We can further drop the correlation between the channel gains of the first receiver and the first antenna of the second receiver, as the capacity of a BC only depends on the marginal distributions of the received signals. 
Thus for the MIMO BC described above under no CSIT, we have
\begin{align}
n & \left( m R_1 + R_2 - \epsilon_n \right) \nonumber \\
& \leq \sum_{j=1}^{m}{\left\{ I\left( W_1; Y_{d_1'}^n | \Hmimo^n\right) \right\}} + I\left( W_2; Y_{d_2^\prime}^n | \Hmimo^n\right) \nonumber \\
& = \sum_{j=1}^{m}{\left\{ I\left( W_1; Y_{v_j}^n | \Hmimo^n\right) \right\}} + I\left( W_2; Y_{d_2^\prime}^n | W_1, \Hmimo^n\right) \nonumber \\
& = \sum_{j=1}^{m}{\left\{ h\left( Y_{v_j}^n | \Hmimo^n\right) - h\left( Y_{v_j}^n | W_1, \Hmimo^n\right) \right\}} \nonumber \\
& \quad + h\left( Y_{d_2^\prime}^n | W_1, \Hmimo^n\right) - h\left( Y_{d_2^\prime}^n | W_1, W_2, \Hmimo^n\right) \nonumber \\
& = \sum_{j=1}^{m}{\left\{ h\left( Y_{v_j}^n | \Hmimo^n\right) - h\left( Y_{v_j}^n | W_1, \Hmimo^n\right) \right\}} \nonumber \\
& \quad + h\left( Y_{d_2^\prime}^n | W_1, \Hmimo^n\right) - h\left( Z_{v_1}^n, \ldots, Z_{v_m}^n | \Hmimo^n\right) \nonumber \\
& \eqnum \sum_{j=1}^{m}{\left\{h\left( Y_{v_j}^n | \Hmimo ^n\right) - h\left( Z_{v_j}^n | \Hmimo^n\right)\right\}} \nonumber \\
& \quad - \left[ \sum_{j=2}^{m}{I\left( Y_{v_1}^n, \ldots, Y_{v_{j-1}}^n ; Y_{v_{j}}^n | W_1, \Hmimo^n \right)} \right] \nonumber \\
& \leq \sum_{j=1}^{m}{\left\{h\left( Y_{v_j}^n | \Hmimo ^n\right) - h\left( Z_{v_j}^n | \Hmimo^n\right)\right\}} \nonumber \\
& \leq m n \left( \tfrac12\log P + o\left( \log P \right)\right),
\end{align} \rescnt
where \cnt follows since 
\aln{
& h\left( Y_{d_2^\prime}^n | W_1, \Hmimo^n\right) - \sum_{j=1}^{m}{ h\left( Y_{v_j}^n | W_1, \Hmimo^n\right) }  \\
& = \sum_{j=2}^{m}{ h\left( Y_{v_j}^n | Y_{v_1}^n,..., Y_{v_{j-1}}^n, W_1, \Hmimo^n\right) - h\left( Y_{v_j}^n | W_1, \Hmimo^n\right) } \\
& = \sum_{j=2}^{m}{ I\left(  Y_{v_1}^n,..., Y_{v_{j-1}}^n; Y_{v_j}^n | W_1, \Hmimo^n\right).  }
}
Therefore, we conclude that 
\begin{align}
mD_1+D_2 \leq m,
\end{align}
which completes the proof of Lemma~\ref{bclem2}. 
\end{proof}

Now once again, consider the network of Fig.~\ref{Fig:3D1D2}. 
In this network, $v_5$ is a $3$-bottleneck node for $d_1$. 
Thus for this network, using Theorem~\ref{bottlethm}, we have
\begin{equation}
\label{eq:DoF3D1D2}
\left\{ \begin{array}{ll}
\vspace{1mm} 0 \leq D_i \leq 1, & i = 1,2, \\
3 D_1 + D_2 \leq 3. &
\end{array} \right.
\end{equation}
In Section~\ref{Section:Examples}, we provided the achievability proof of corner point $\left( D_1, D_2 \right) = \left( 2/3, 1 \right)$. As a result, the outer-bound provided by Theorem~\ref{bottlethm} (alongside individual bounds) completely characterizes the achievable DoF region in this case.


\section{Proof of Theorem~\ref{mainthm}}
\label{Section:Proof}


In this section, we describe the proof of Theorem~\ref{mainthm}. 
In essence, we show that the example considered in Section~\ref{bottleexsec} can be generalized to a class of networks that contain bottleneck nodes whose corresponding outer bounds can be achieved.

\begin{figure}[ht]
\centering
\includegraphics[height = 4cm]{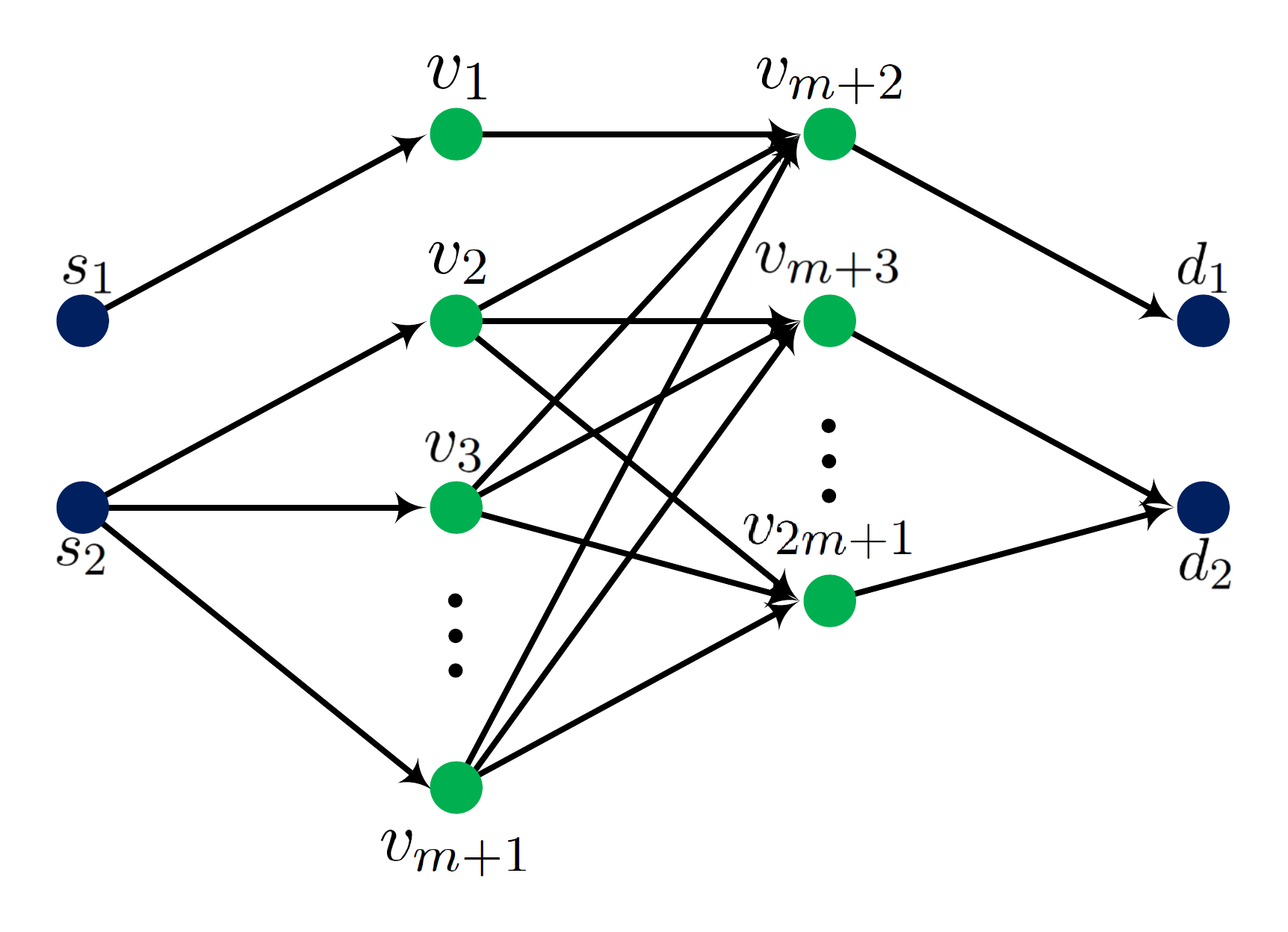}
\caption{In this example, we show that we can achieve corner point $\left( D_1, D_2 \right) = \left( (m-1)/m, 1\right)$.}\label{Fig:mD1D2}
\end{figure}

First we consider the network illustrated in Fig.~\ref{Fig:mD1D2}.
In this network, $v_{m+2}$ is an $m$-bottleneck node for $d_1$, and there is no bottleneck node for $d_2$.
We show that we can achieve corner point $\left( D_1, D_2 \right) = \left( (m-1)/m, 1\right)$. The achievability strategy is a generalization of the strategy presented for the network of Fig.~\ref{Fig:3D1D2}, and uses $m$ time steps.
As in that case, the transmission scheme for the first and third hops is straightforward and we only focus on the second hop. 

\noindent {\bf Transmission strategy for the intermediate problem:} The transmission strategy has $m$ time slots. During the first time slot, relay $v_1$ remains silent and relay $v_j$ sends out symbol $b_{j-1}$ intended for destination $d_2$, $j=2,3,\ldots,m+1$. Ignoring the noise terms, relay $v_j$ obtains a linear combination of the symbols intended for destination $d_2$, $L_{j-m-1}\left( \vec{b} \right)$, $j = m+2, \ldots,2m+1$.

\begin{figure}[ht]
\centering
\includegraphics[width = \linewidth]{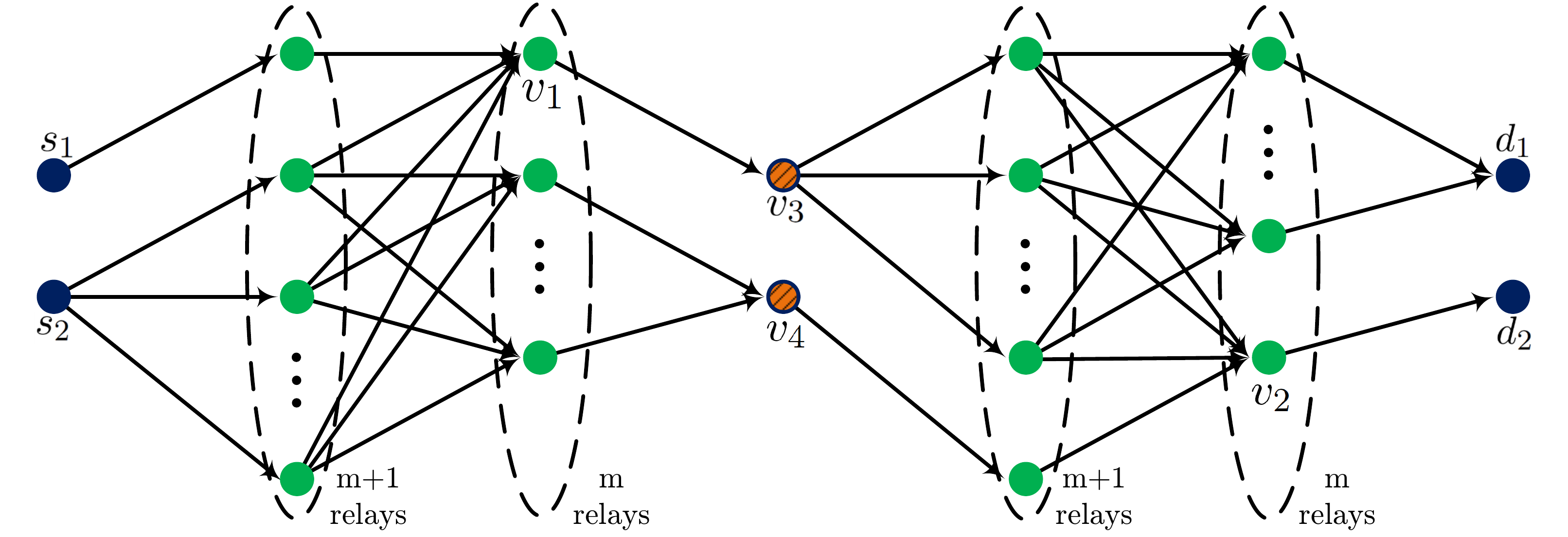}
\caption{Relay $v_1$ is an $m$-bottleneck node for $d_1$ and relay $v_2$ is an $m$-bottleneck node for $d_2$.} \label{Fig:TwoBounds}
\end{figure}

At this point, using the delayed knowledge of the channel state information, relay $v_2$ can (approximately) reconstruct $L_1(\vec{b})$. During the second time slot, relay $v_1$ sends out $a_1$, relay $v_3$ sends out $L_1(\vec{b})$ (normalized to meet the power constraint), and relays $v_3,\ldots, v_{m+1}$ remain silent. This way, $v_{m+2}$ obtains a linear combination of $a_1$ and $L_1(\vec{b})$ denoted by $L_{m+1} (a_1, L_1(\vec{b}) )$. Note that $v_{m+2}$ already has access to $L_1(\vec{b})$ and thus can recover $a_1$. Also, $L_1(\vec{b})$ becomes available to $v_j$ for $j=2,3,\ldots,m+1$.

During time slot $\ell$, $\ell = 3, \ldots, m$, relay $v_1$ sends out $a_{\ell-1}$ and relays $v_2, v_3, \ldots, v_{m+1}$ remain silent. Note that with this strategy, $v_{m+2}$ obtains $a_1, a_2, \ldots, a_{m-1}$, and relays $v_2, v_3, \ldots, v_{m+1}$ (with probability 1) obtain $m$ linearly independent combinations of $b_1, b_2, \ldots, b_m$. Then the task for the third hop is to simply deliver $a_1, ..., a_{m-1}$ to $d_1$ and the $m$ linearly independent combinations of $b_1, ..., b_m$ to $d_2$.


Since we have matching inner and outer bounds, we conclude that for the network in Fig.~\ref{Fig:mD1D2}, the sum DoF are $\dE = 1 + (m-1)/m = 2 - 1/m$, for $m \in \{1,2,...\}$.
Notice that this corresponds to half of the values in the set $\S$ in \eref{Seq}.
To obtain the remaining values in $\S$, we need a class of networks that contain both a bottleneck node for $d_1$ and a bottleneck node for $d_2$.

Consider the network depicted in Fig.~\ref{Fig:TwoBounds}. 
For simplicity of notation, we have only labeled a few relays in this network. 
We claim that for this network $\dE = 2m/(m+1)$, $m \in \{1,2,...\}$.
First, we prove the converse. 
It is straightforward to verify that relay $v_1$ is an $m$-bottleneck node for $d_1$ and relay $v_2$ is an $m$-bottleneck node for $d_2$. 
Thus, from Theorem~\ref{bottlethm}, we have 
\al{ 
m D_{i} + D_{\bar{i}} \leq m, \qquad i=1,2.
}
To prove that the outer-bounds are tight, it suffices to prove the achievability of corner point $\left( D_1, D_2 \right) = \left( m/(m+1), m/(m+1) \right)$. 

\vspace{2mm}
\noindent {\bf Transmission strategy:} The goal is to deliver $m$ symbols to each destination during $m+1$ time slots. Denote the symbols intended for $d_1$ by $a_i$'s and the symbols intended for $d_2$ by $b_i$'s, $i=1,2,\ldots,m$. 
We point out that the network in Fig.~\ref{Fig:TwoBounds} can be seen as a concatenation of the network in Fig.~\ref{Fig:mD1D2} with flipped copy of itself.
Hence, we will describe the achievability in terms of each of the two subnetworks.
We first describe how to deliver $a_i$'s to relay $v_3$ and $b_i$'s to relay $v_4$. 
Then, the goal becomes for relay $v_3$ to deliver $a_i$'s to $d_1$ and for relay $v_4$ to deliver $b_i$'s to $d_2$, $i=1,2,\ldots,m$. 
Since the two subnetworks are essentially identical, 
we only need to show that we can deliver $a_i$'s to relay $v_3$ and $b_i$'s to relay $v_4$ during $m+1$ time slots. 
Then, the relays in the second subnetwork 
will implement a similar strategy to that of the nodes in the first subnetwork. 

Since the first subnetwork is identical to the network of Fig.~\ref{Fig:mD1D2}, by using the same strategy, during $m$ time slots we can deliver $m-1$ symbols to $v_3$ and $m$ symbols to $v_4$. 
During the last time slot, \emph{i.e.} time slot $m+1$, source $s_2$ remains silent, and source $s_1$ sends out one more symbol, $a_m$, to relay $v_3$. 
This way, we successfully deliver $a_i$'s to relay $v_3$ and $b_i$'s to relay $v_4$
during $m+1$ time slots, $i=1,2,\ldots,m$. 
Repeating the same strategy over the second subnetwork, each destination can decode its $m$ symbols over $m+1$ time steps, and we conclude 
that $\dE = 2m/(m+1) = 2-2/(m+1)$.
This completes the proof of Theorem~\ref{mainthm}. 

\section{Discussion}
\label{Section:Discussion}


In this paper, we introduced a new technique to derive outer bounds on the DoF of two-unicast wireless networks with delayed CSIT, and we presented several transmission strategies that can achieve these outer bounds.
The presented transmission strategies achieve the optimal DoF in a finite number of time slots. In this section, we discuss two follow-up questions to our main results:
\begin{enumerate}
\item Do bounds of the form $m D_i + D_{\bar{i}} \leq m$ for $m \geq 1$ suffice to characterize the DoF region of the two-unicast wireless networks with delayed CSIT?
\item Can we achieve the optimal DoF region of a two-unicast wireless networks with delayed CSIT in a finite and bounded number of time slots?
\end{enumerate}

\begin{figure}[ht]
\centering
\includegraphics[height = 3cm]{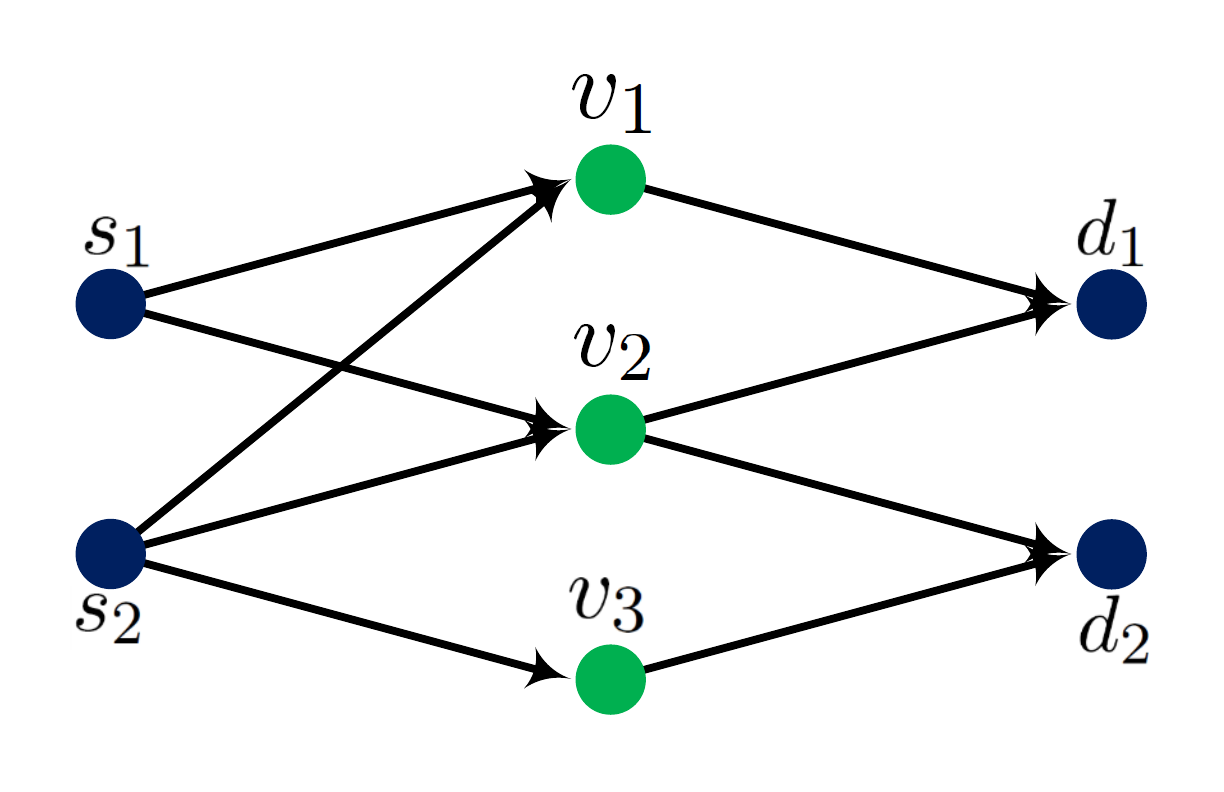}
\caption{The outer-bound provided by Theorem~\ref{bottlethm} does not describe the DoF region of this network. Moreover, the achievability strategy for the corner points of the DoF region does not have a finite number of time slots.}\label{Fig:D1D2onehalf}
\end{figure}

As it turns out, the answers to the questions posed above are both negative. To provide some insights, we consider the network depicted in Fig.~\ref{Fig:D1D2onehalf}. Under instantaneous CSIT assumption, the DoF region of this network is derived in~\cite{dof2unicastfull} and is given by
\begin{equation}
\label{eq:DoFRegionIlan}
\left\{ \begin{array}{ll}
\vspace{1mm} 0 \leq D_i \leq 1, & i = 1,2, \\
D_1 + D_2 \leq \frac{3}{2}. &
\end{array} \right.
\end{equation}
Interestingly, under the delayed CSIT assumption, we can still achieve this region. 
However, the network in Fig.~\ref{Fig:D1D2onehalf} contains no bottleneck nodes.
Moreover, it can be verified that the region in \eref{eq:DoFRegionIlan} cannot be obtained 
from bounds of the form $m D_i + D_{\bar{i}} \leq m$ for $m \geq 1$. 

Next, we briefly describe the achievability strategy for corner point $\left( D_1, D_2 \right) = \left( 1, 1/2 \right)$. The achievability strategy goes over $2m+1$ time slots and upon completion of the transmission, we achieve
\begin{align}
\left( D_1, D_2 \right) = \left( \frac{2m}{2m+1}, \frac{k}{2m+1} \right),
\end{align}
where $m$ is an arbitrarily chosen parameter.
Thus, as the number of time slots $m$ goes to infinity, we achieve arbitrarily close to the corner point $\left( D_1, D_2 \right) = \left( 1, 0.5 \right)$.

\begin{figure*}[t]
\centering
\includegraphics[width = 17cm]{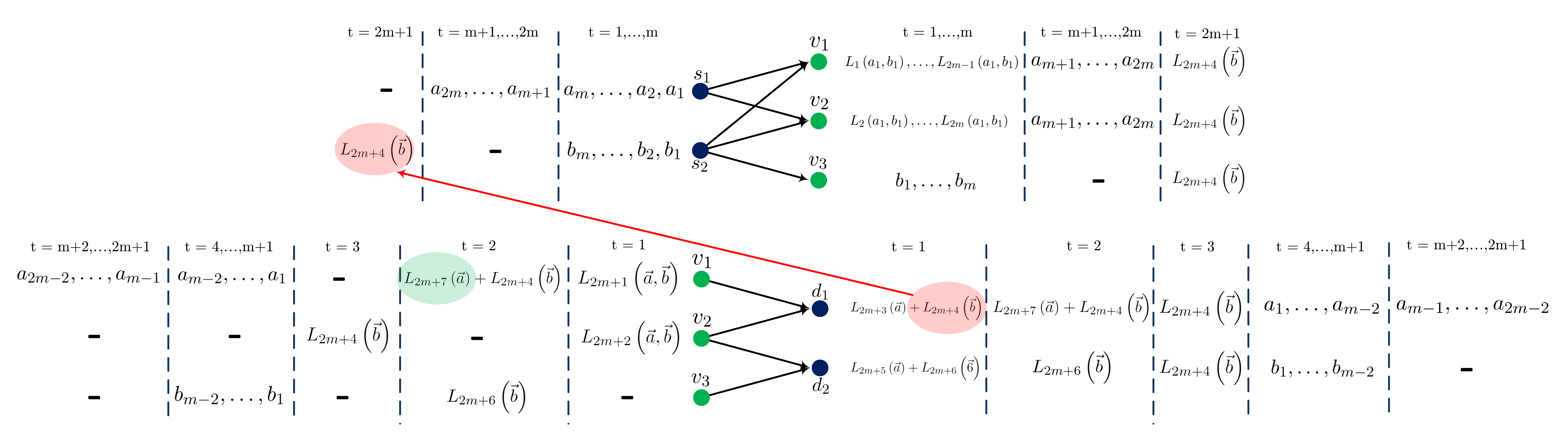}
\caption{Achievability strategy for corner point $\left( D_1, D_2 \right) = \left( 1, 1/2 \right)$ of the DoF region of the network of Fig.~\ref{Fig:D1D2onehalf}. The achievability strategy uses $2m+1$ time slots and as $m \to \infty$, we achieve the desired corner point.} \label{Fig:D1D2onehalf-Cornerpoint}
\end{figure*}

The transmission strategy is illustrated in Fig.~\ref{Fig:D1D2onehalf-Cornerpoint}. We highlight the important aspects of this strategy here. First we note that by interleaving different blocks, we encode such that the first $2m$ time slots of the first hop occur before the first time slot of the second hop. This way, there will be no issues regarding causality in the network.

For the first hop, the communication during the first $2m$ time slots is straightforward. In the second hop during the first time slot, relays $v_1$ and $v_2$ create random linear combinations of all the signals they received during the first $2m$ time slots of the first hop and send them out. Destination one obtains $L_{2m+3}\left(\vec{a}\right) + L_{2m+4}(\vec{b})$, and destination two obtains $L_{2m+5}\left(\vec{a}\right) + L_{2m+6}(\vec{b})$. Our goal is to deliver $L_{2m+4}(\vec{b})$ to both receivers. Relay $v_3$ can reconstruct $L_{2m+4}(\vec{b})$, however, there is no link form $v_3$ to destination one. As a result, during the final time slot, the second source sends out $L_{2m+4}(\vec{b})$ and this signal becomes available to all receivers (see Fig.~\ref{Fig:D1D2onehalf-Cornerpoint} where $L_{2m+4}(\vec{b})$ is highlighted by a red oval). 

The key idea for the achievability would be the observation that relay $v_1$ can combine its previous observations in a way that $b_i$'s form $L_{2m+4}(\vec{b})$, $i=1,2,\ldots,m$. This way, during the first two time slots, the interference at destination one would be the same. Thus, if we provide $L_{2m+4}(\vec{b})$ to destination one, it can recover $L_{2m+3}\left(\vec{a}\right)$ and $L_{2m+7}\left(\vec{a}\right)$. Finally, we note that $L_{2m+6}(\vec{b})$ is linear combination of $b_i$'s that destination two obtains during the first time slot.

Upon completion of the transmission strategy, destination one has access to 
\begin{align}
a_1, a_2, \ldots, a_{2m-2}, L_{2m+3}\left(\vec{a}\right), L_{2m+7}\left(\vec{a}\right).
\end{align}
Hence, receiver one has enough equations to recover its intended symbols. Similarly, destination two has access to 
\begin{align}
b_1, b_2, \ldots, a_{m-2}, L_{2m+4}(\vec{b}), L_{2m+6}(\vec{b}),
\end{align}
which allows destination two to recover its intended symbols.

%
%
%


\bibliographystyle{ieeetr}
\bibliography{bib_TwoUnicastDelayed}

\end{document}